\documentclass[prd,showpacs,superscriptaddress,floatfix,nofootinbib,10pt]{revtex4-2}
\usepackage{times}
\usepackage{amssymb,amsbsy,amsmath,amsfonts}
\usepackage{graphicx}
\usepackage{float}
\usepackage{color}
\usepackage{orcidlink}
\usepackage{morefloats}
\usepackage{rotating}
\usepackage{srcltx}
\usepackage{slashed}
\usepackage{multirow}
\usepackage{verbatim}
\usepackage{hyperref}
\usepackage{tabularx}
\usepackage{bm}
\allowdisplaybreaks[4]

\usepackage{bbding}
\usepackage{threeparttable}

\usepackage{mathrsfs}

\newcommand{\PreserveBackslash}[1]{\let\temp=\\#1\let\\=\temp}
\newcolumntype{C}[1]{>{\PreserveBackslash\centering}p{#1}}
\newcolumntype{R}[1]{>{\PreserveBackslash\raggedleft}p{#1}}
\newcolumntype{L}[1]{>{\PreserveBackslash\raggedright}p{#1}}

\begin{document}

\title{$K\,X(3872)$ interaction and correlation function}

\author{Jing Song\orcidlink{0000-0003-3789-7504}}
\email[]{Song-Jing@buaa.edu.cn}
\affiliation{Center for Theoretical Physics, School of Physics and Optoelectronic Engineering, Hainan University, Haikou 570228, China}
\affiliation{School of Physics, Beihang University, Beijing, 102206, China}
\affiliation{Departamento de Física Teórica and IFIC, Centro Mixto Universidad de Valencia-CSIC Institutos de Investigación de Paterna, 46071 Valencia, Spain}

\author{Pedro Brandão\orcidlink{0009-0003-1630-9336}}
\email[]{pedro.brandao@ufba.br}
\affiliation{Instituto de Física, Universidade Federal da Bahia, Campus Ondina, Salvador, Bahia 40170-115, Brazil}
\affiliation{Departamento de Física Teórica and IFIC, Centro Mixto Universidad de Valencia-CSIC Institutos de Investigación de Paterna, 46071 Valencia, Spain}

\author{Eulogio Oset\orcidlink{0000-0002-4462-7919}}%
\email[]{oset@ific.uv.es}
\affiliation{Departamento de Física Teórica and IFIC, Centro Mixto Universidad de Valencia-CSIC Institutos de Investigación de Paterna, 46071 Valencia, Spain}	
\affiliation{Department of Physics, Guangxi Normal University, Guilin 541004, China}


\begin{abstract}
We investigate the $K X(3872)$ interaction and the corresponding correlation function, assuming the $X(3872)$ to be a molecular state of $D \bar D^*$ and $D^* \bar D$ with isospin $I=0$ and positive $C$-parity.
The interaction is treated within the fixed-center approximation (FCA) to the Faddeev equations, in which the $X(3872)$ is taken as a cluster of its constituents and the kaon interacts with the $D^*$ and $D$ components.
The three-body scattering amplitude is evaluated using the Fixed Center Approximation (FCA) to the Faddeev equations, improved by taking the  FCA amplitude as an optical potential which is later unitarized by means of the Lippmann–Schwinger equation. 
We find a narrow resonant structure about 50~MeV below the $K^+ X(3872)$ threshold with a width of approximately 1~MeV, and determine the $K X(3872)$ scattering length $a = (0.39 - i\,0.00)$~fm and effective range $r_0 = (1.16 - i\,1.66)$~fm.
The corresponding correlation function is evaluated and shows a clear deviation from unity at low momenta, characteristic of a strongly attractive interaction leading to a bound state.
These predictions are tied to the molecular nature of the $X(3872)$ and can be measured experimentally via measurements of the $K X(3872)$ correlation function and three-body invariant mass distributions.
\end{abstract}

\maketitle

\section{Introduction}\label{sec:Intr}

The investigation of correlation functions in hadron dynamics has entered a new phase with the study of correlation functions for three-body systems involving a stable particle and a resonance.
A particularly promising case is the measurement of the $p f_1(1285)$ correlation function by the ALICE collaboration~\cite{Serksnyte:2026,ALICE:2024rjz}, which stimulated intense theoretical efforts to understand the nature of resonances from their interactions with external probes.
The $f_1(1285)$ is widely described as a molecular state dynamically generated from the $K \bar K^*$ and $\bar K K^*$ interactions in isospin $I=0$~\cite{Lutz:2003fm,Roca:2005nm,Garcia-Recio:2010enl,Guo:2017jvc}, a picture that  has been  established by experimental observations of its decays and production reactions~\cite{Aceti:2015zva,Aceti:2015pma,Xie:2015wja}.
A similar molecular interpretation is given to the well-known exotic state $X(3872)$, which is considered as a shallow $D \bar D^*$ bound state with $J^{PC}=1^{++}$~\cite{Tornqvist:2004qy,Braaten:2007dw,Gamermann:2009uq,Harada:2010bs,Kamiya:2022thy,Wang:2022xga,Song:2023pdq,Ji:2025hjw}.
The small mass difference between the $X(3872)$ and the $D^0 \bar D^{*0}$ threshold, the isospin breaking in its decays, and the detailed line shapes observed in the $J/\psi \pi^+\pi^-$ and $D^0\bar D^0\pi^0$ channels all strongly support this molecular picture~\cite{Belle:2011vlx,ATLAS:2016kwu,Guo:2017jvc,Kamiya:2022thy,Song:2023pdq}.
Alternative interpretations, such as a compact tetraquark~\cite{Maiani:2004vq}, have been proposed. A recent study of the BESIII data on $e^+e^-\to \gamma(D^0\bar D^0\pi^0)J/\psi\pi^+\pi^-$, and LHCb data on $B^+\to K^+(J/\psi\pi^+\pi^-)$, with full consideration of three-body effects, respecting analyticity and unitarity, provides strong support for the molecular picture~\cite{Ji:2025hjw}.

The interaction of a kaon with the $X(3872)$ is analogous to the scattering of an external particle with a two-body bound cluster, similar to the extensively investigated $\bar{K} d$ system~\cite{Toker:1981zh,Kamalov:2000iy,Bahaoui:2003xb,Meissner:2005bz,Mizutani:2012gy}.
The Fixed Center Approximation (FCA) to the Faddeev equations~\cite{Kamalov:2000iy,Deloff:1999gc,Foldy:1945zz} provides a practical  framework for studying such a system,  assuming the $X(3872)$ to be a $D\bar D^*$ cluster in the scattering process.
A known shortcoming of the traditional FCA is the violation of elastic unitarity at the threshold of the external particle and the cluster, a problem that was initially circumvented by introducing a multiplicative factor in the study of $p f_1(1285)$~\cite{Encarnacion:2025lyf} and  it was formally solved in Ref.~\cite{Ikeno:2025bsx} by relating the three-body amplitude to the solution of a Lippmann-Schwinger equation with an optical potential.
In this approach, the optical potential is constructed from the elementary two-body scattering matrices of the kaon with the $D$ and $\bar D^*$ constituents of the $X(3872)$, and the scattering is resummed within a framework equivalent to the Lippmann-Schwinger equation that guarantees elastic unitarity.
This modified unitary FCA has been successfully applied to predict bound states and correlation functions in several systems, including $n \bar D_{s0}^*(2317)$~\cite{Ikeno:2025bsx}, $n \bar D_{s1}$~\cite{Agatao:2025ckp}, $K^* D^* K^*$~\cite{Jia:2025obs}, $K f_1(1285)$~\cite{Jia:2026dpl}, $K D_{s0}^*(2317)$~\cite{Jia:2026iqo}, and $\pi^0 f_1(1285),~\eta f_1(1285)$~\cite{Jia:2026ewk}.

{In the present work,} we extend these theoretical developments to the $K X(3872)$ system, assuming the $X(3872)$ as a $D \bar D^*$ molecular state.
We evaluate the $K X(3872)$ scattering amplitude using the unitarized FCA, determine the threshold scattering parameters (scattering length and effective range), and evaluate the corresponding correlation function within the formalism of Ref.~\cite{Ikeno:2025bsx}.
A bound state is also predicted in this sector, and its existence is closely related  to the composite nature of the $X(3872)$.
The richness of the system stems from the fact that the kaon can interact with both the $D$ and $ D^*$ components, giving rise to a coupled-channel dynamics that can be directly investigated via femtoscopic measurements.
Given that the experiment already has the capability to detect the $X(3872)$ through its $J/\psi \pi^+ \pi^-$ decay~\cite{ATLAS:2016kwu}, the measurement of the $K X(3872)$ correlation function is experimentally feasible and would provide novel, complementary information on the nature of the $X(3872)$.
The observables derived here can also be utilized in the inverse method of Refs.~\cite{Ikeno:2023ojl,Albaladejo:2023wmv} to extract the interaction parameters directly from future experimental data, serving as a critical measure of the molecular picture of the $X(3872)$ and the validity of the three-body theoretical framework.
{We should note that the present idea was already developed in Ref.~\cite{Ren:2018pcd}, using the same framework without the unitarization. It is also remarkable that in that work a bound state was also found at about the same energy than we find here, which indicates the fairness of the standard FCA to make predictions of bound states. Yet, the threshold region and the correlation function, which we study here for the first time, requires the unitary  formalism that we present here. }

\section{Formalism}\label{sec:form}
The $X(3872)$ is assumed to be a $D\bar{D}^*- D^*\bar{D}$ 
molecular state, with quantum number $I^{G}(J^{PC}) = 0^{+}(1^{++})$~\cite{Gamermann:2009uq,Song:2023pdq}.
We adopt the isospin conventions
\begin{equation}
	\begin{pmatrix} D^{+} \\ -D^0 \end{pmatrix},\quad
	\begin{pmatrix} \bar D^{0} \\ D^- \end{pmatrix},\quad
	\begin{pmatrix} D^{*+} \\ -D^{*0} \end{pmatrix},\quad
	\begin{pmatrix} \bar D^{*0} \\ D^{*-} \end{pmatrix},
\end{equation}
which lead to the following $I=0$ combinations
\begin{align}
	| D \bar D^*, I=0 \rangle &= \frac{1}{\sqrt{2}}\big( D^{+} D^{*-} + D^{0}\bar D^{*0} \big), \label{eq:I0a} \\
	| D^{*}\bar D, I=0 \rangle &= \frac{1}{\sqrt{2}}\big( D^{*+} D^- + D^{*0}\bar D^0 \big). \label{eq:I0b}
\end{align}
{In our formalism, the C parity acts as
$C |D\rangle = |\bar{D}\rangle$,  
$C |D^{*}\rangle = - |\bar{D}^{*}\rangle$.
}
The physical $X(3872)$ state is then expressed as the symmetric combination
\begin{equation}
	X(3872) = \frac{1}{\sqrt{2}}\Big( |D^{*}\bar D\rangle_{I=0} - |D\bar D^{*}\rangle_{I=0} \Big),
	\label{eq:Xwave}
\end{equation}
which has positive $C-$parity.

\begin{figure}[t]
	\begin{center}
		\includegraphics[width=0.68\textwidth]{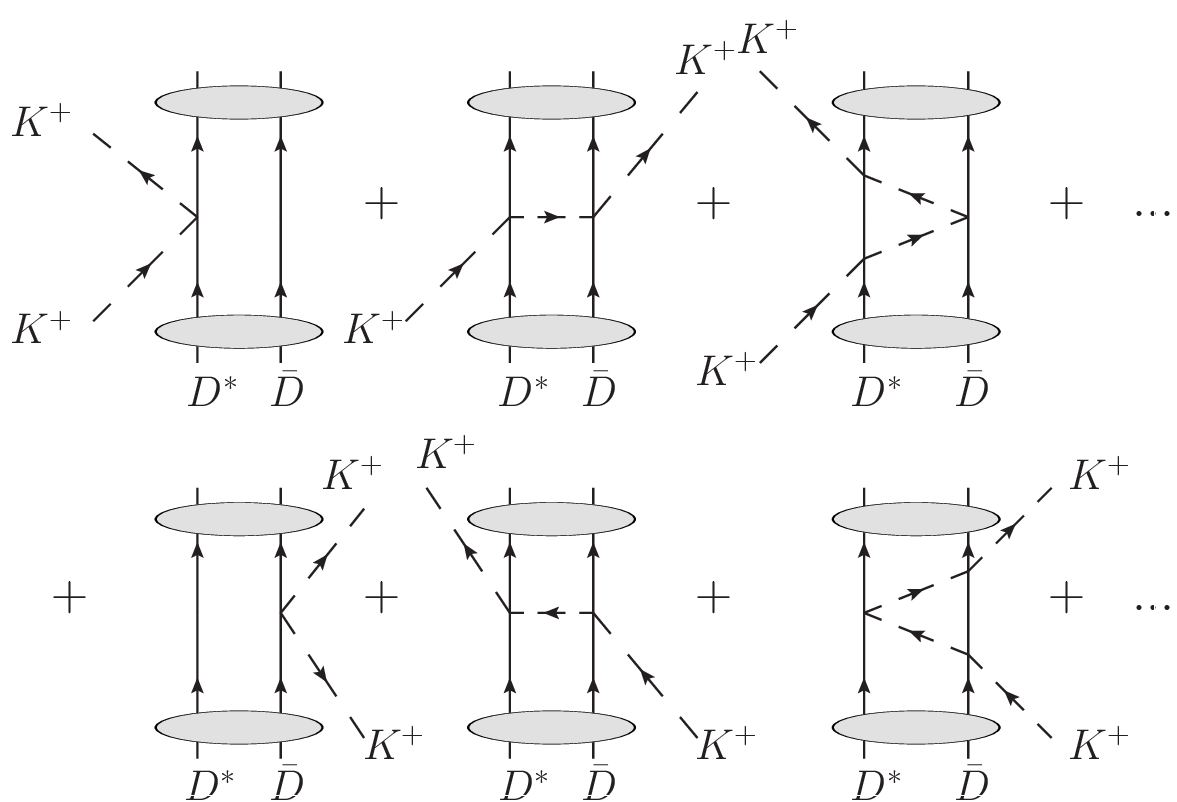}
	\end{center}
	\vspace{-0.5cm}
	\caption{Diagrams entering the ordinary FCA approach for $K^+$ interacting with the cluster of $D^*\bar D$.}
	\label{Fig1}
\end{figure}

We follow closely the formalism of Refs.~\cite{Ikeno:2025bsx,Agatao:2025ckp,Jia:2025obs}, in particular the one of Ref.~\cite{Jia:2025obs} which involves only mesons.  In the
conventional FCA approach {we have two sets of diagrams,}
those of Fig.~\ref{Fig1}, which correspond to $K^+$ interacting with the $D \bar D^* (I=0)$ component of the $X(3872)$ state,
and we also have to consider the similar diagrams for the interaction of $K^+$ with the $D^*\bar D$ component.
For reasons of normalization, we refer the amplitude to the $K^+ X(3872)$ system with the standard normalization of the $K^+$ and $X(3872)$ fields, and we write the individual amplitudes as~\cite{Roca:2010tf}
\begin{equation}
	\tilde{t}_1 = \frac{M_c}{M_{D^*}}\, t_1 \,\,,
	\qquad
	\tilde{t}_2 = \frac{M_c}{M_{D}}\, t_2 \,\,,
	\label{eq:1}
\end{equation}
where $M_c$ is the mass of the cluster $X(3872)$ {(we take $M_c=m_{X}=3871.65$~MeV)} and $t_1, t_2$ are the scattering matrices for the interaction of $K^+$ with $D^*$ and $\bar D$ respectively.
Considering that the system $D^* \bar D$ is in $I=0$ in the cluster we find that \cite{Ikeno:2025bsx}
\begin{equation}
	\begin{split}
		t_1 &= \frac{3}{4} \,t_{K D^*}^{I=1} + \frac{1}{4} \,t_{K D^*}^{I=0},\\[1.5mm]
		t_2 &= \frac{3}{4}\, t_{K \bar D}^{I=1} + \frac{1}{4} \,t_{K \bar D}^{I=0}.
	\end{split}
	\label{eq:2}
\end{equation}
Then we define $\tilde{T}_{ij}$ as the sum of the diagrams where the $K^+$ interacts initially with particle $i$ of the cluster and finishes in particle $j$ of the cluster in the sum of diagrams of Fig.~\ref{Fig1}.
One obtains
\begin{equation}
    \tilde{T} = \begin{pmatrix} \tilde{T}_{11} & \tilde{T}_{12} \\[1mm] 
	\tilde{T}_{21} & \tilde{T}_{22} \end{pmatrix},
\end{equation}
with
\begin{equation}
	\begin{split}
		& \tilde{T}_{11} = \frac{\tilde{t}_1}{1 - \tilde{t}_1 \,\tilde{t}_2 \,G_0^2}, \\[1.5mm]
		& \tilde{T}_{22} = \frac{\tilde{t}_2}{1 - \tilde{t}_1\, \tilde{t}_2 \, G_0^2}, \\[1.5mm]
		& \tilde{T}_{12} = \tilde{T}_{21} = \frac{\tilde{t}_1 \,\tilde{t}_2 \,G_0}{1 - \tilde{t}_1 \,\tilde{t}_2 \, G_0^2},
	\end{split}
	\label{eq:3}
\end{equation}
where
\begin{align}\label{eq:4}
	G_0(\sqrt{s}) &= \int \frac{\mathrm{d}^3 q}{(2\pi)^3} \,
		\frac{F_c(q)}{\sqrt{s} - \omega_{K}(\vec q\,) - \omega_c(\vec q \,) + i \epsilon} \; \dfrac{1}{2\, \omega_{K}(\vec q\,)} \nonumber \\[1.5mm]
	&\times \dfrac{1}{2\, \omega_{c}(\vec q\,)}\; \Theta\left(q_{\mathrm{max}}^{(1)} - q_1^*\right) \, \Theta\left(q_{\mathrm{max}}^{(2)} - q_2^*\right),
\end{align}
with $\omega_{K}(\vec q\,) = \sqrt{M_{K^+}^2 + \vec{q}^{\;2}}$ and $\omega_c(\vec q\,) = \sqrt{M_c^2 + \vec{q}^{\;2}}$.
In Eq.~\eqref{eq:4}, $F_c(q)$ is the form factor of the cluster, which considering the wave function in momentum space for the cluster coming from our approach \cite{Gamermann:2009uq,Song:2023pdq}
{
\begin{equation}
	\Psi(p)=g\;\frac{\Theta\left(q_{\mathrm{max}}-|\vec{p}\,|\right)}
	{M_c - \omega_{D^*}(\vec{p}\,) - \omega_{\bar D}(\vec p\,)},
	\label{eq:5}
\end{equation}
}
\noindent with $g$ the coupling of the state to the $ D^* \bar D$ component, can be written as
\begin{equation}
	\begin{aligned}
		F_c(q) &= \frac{F(q)}{N}, \\[1.5mm]
		F(q) &= \int\limits_{\substack{|\vec{p}\,| < q_{\mathrm{max}} \\ |\vec{p} - \vec{q}\,| < q_{\mathrm{max}}}} 
		\frac{\mathrm{d}^3 p}{(2\pi)^3} \,
		\frac{1}{M_c - \omega_{D^*}(\vec{p}\,) - \omega_{\bar D}(\vec{p}\,)} \\
		&\quad \times \frac{1}{M_c - \omega_{D^*}(\vec{p} - \vec{q}\,) - 
			\omega_{\bar D}(\vec{p} -\vec{q}\,)}, \\[2mm]
		N &= F(0) \\
		&= \int\limits_{|\vec{p}\,| < q_{\mathrm{max}}} \frac{\mathrm{d}^3 p}{(2\pi)^3} \,
		\left[ \frac{1}{M_c - \omega_{D^*}(\vec{p}\,) - \omega_{\bar D}(\vec{p}\,)} \right]^2.
	\end{aligned}
	\label{eq:6}
\end{equation}
In Eq.~\eqref{eq:5}, $q_{\mathrm{max}}$ is the cutoff in the meson-meson loop functions, which arises naturally from assuming a potential of the type $V\left(\vec{q},\vec{q}\,'\right)=V\;\Theta\left(q_{\mathrm{max}}-|\vec{q}\,|\right)\;\Theta\left(q_{\mathrm{max}}-|\vec{q}\,'|\right)$, leading to the same factorization in the scattering matrix $t\left(\vec{q},\vec{q}\,'\right)=t\;\Theta\left(q_{\mathrm{max}}-|\vec{q}\,|\right)\;\Theta\left(q_{\mathrm{max}}-|\vec{q}\,'|\right)$. 
{We have $q_{\mathrm{max}}=288$~MeV which is determined from a single-channel $DD^*$ calculation and reproduces the mass of the $X(3872)$ accurately~\cite{Song:2023pdq}.}
%
In Eq.~\eqref{eq:4}, $q_{\mathrm{max}}^{(1)}$ and $q_{\mathrm{max}}^{(2)}$ refer to the cutoffs associated with the $K^+D^*$ and $K^+\bar{D}$ scattering matrices, respectively, whose values are found in the Appendix A.
$q_1^*$ and $q_2^*$ are the momenta of the $K^+$ in the rest frames of the corresponding two-body channels, assuming that the momentum transfer is shared equally between 
 initial and final particles of the cluster~\cite{Carrasco:1991we,Boffi:1991nh}.

One has~\cite{Jia:2025obs}
\begin{equation}
	\vec{q}^{\,*}_i = \vec{q}\;\left( 1 - \frac{1}{2}\, \frac{M_{K^+}}{M_{K^+} + M_{i}} \right),
	\label{eq:7}
\end{equation}
with $M_1=M_{D^*}$, $M_2=M_{\bar D}$.

\begin{figure}[t]
	\begin{center}
		\includegraphics[width=0.68\textwidth]{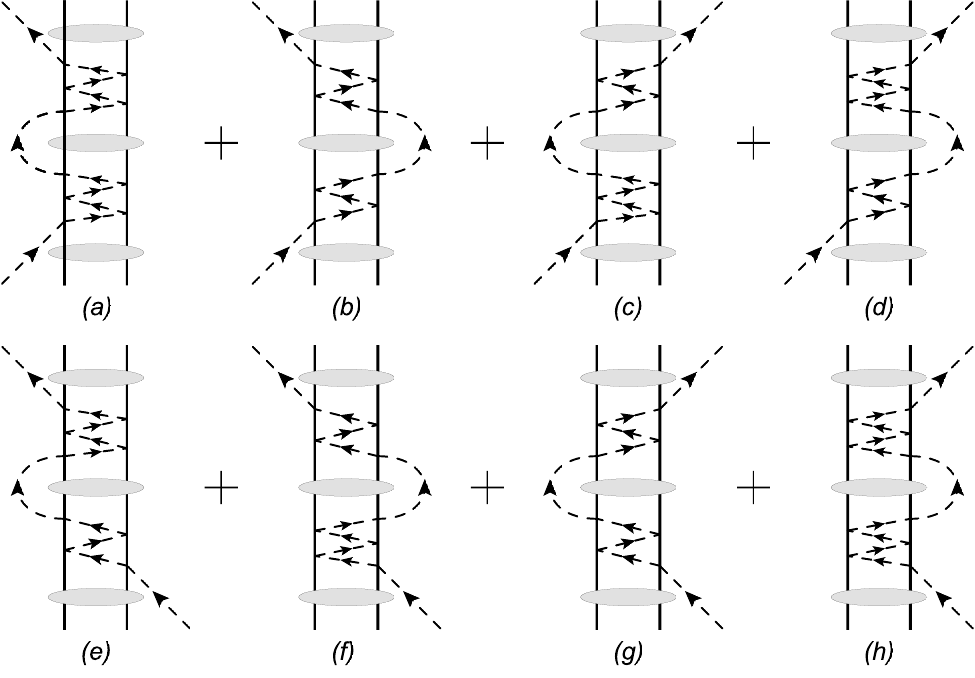}
	\end{center}
	\vspace{-0.5cm}
	\caption{Diagrams considering the elastic propagation of the $K^+$ and the cluster $X(3872)$ as a whole.}
	\label{Fig2}
\end{figure}

The unitarization in the elastic $K^+X(3872)$ channel is achieved by propagating the intermediate $K^+X(3872)$ state, which is equivalent to summing the diagrams of Fig.~\ref{Fig2} and further iteration of these diagrams.
These sums lead to the terms
\begin{equation}
	\tilde{T}' = \begin{pmatrix} \tilde{T}_{11}' & \tilde{T}_{12}' \\[1mm]
	\tilde{T}_{21}' & \tilde{T}_{22}' \end{pmatrix}\,, \\
	\label{eq:8}
\end{equation}
Considering these further iterations of the $K^+X(3872)$ intermediate states in Fig.~\ref{Fig2}, one obtains
\begin{equation}\label{eq:9}
	\tilde{T}'=\left[ 1-\tilde{T}\,G_c \right]^{-1}\;\tilde{T},
\end{equation}
where
\begin{equation}
	G_c=\begin{pmatrix}
	G_c^{(1)} & 0 \\[1mm]
	 0 & G_c^{(2)}
	    \end{pmatrix},
\end{equation}
and
\begin{equation} \label{eq:Floopc}
	\begin{aligned}[b]
		G_c^{(i)}(\sqrt{s}) &= \int \frac{\mathrm{d}^3 q}{(2\pi)^3} \;
		\frac{\left[ F_c^{(i)}(q) \right]^2}{\sqrt{s} - \omega_{K}(\vec q\,) - \omega_c(\vec q\,) + \mathrm{i}\epsilon} \\[1.5mm]
		&\quad \times \dfrac{1}{2\,\omega_{K}(\vec q\,)} \;\dfrac{1}{2\,\omega_{c}(\vec q\,)}\;\Theta\left(q_{\mathrm{max}}^{(i)} - q_i^*\right),
	\end{aligned}
\end{equation}
with \cite{Yamagata-Sekihara:2010kpd}
\begin{equation}
	\begin{aligned}
		F_c^{(1)}(q) &= F_c\left( \frac{M_{\bar D}}{M_{D^*}+M_{\bar D}}\, q\right),\\[2mm]
		F_c^{(2)}(q) &= F_c\left( \frac{M_{D^*}}{M_{D^*}+M_{\bar D}}\, q\right) \,.
	\end{aligned}
\end{equation}
{In order to avoid dealing with the $i$ epsilon in Eq. (\ref{eq:Floopc}), a technical trick is used, subtracting and adding a term that allows to perform the evaluation of the pole contribution analytically~\cite{Albaladejo:2023wmv}. }

One more ingredient is needed, the arguments of the $t_1$ and $t_2$ amplitudes are given by
\begin{equation}
	\begin{split}
		s(K^+ D^*) &= \left(p_{K^+} + p_{D^*}\right)^2 \\[1mm]
		&= M_{K^+}^2 + \left(\xi \, M_{D^*}\right)^2 + 2\,\xi\,M_{D^*}\,q^0,\\[2mm]
		s(K^+ \bar D) &= \left(p_{K^+} + p_{\bar D}\right)^2 \\[1mm]
		&= M_{K^+}^2 + \left(\xi \, M_{\bar D}\right)^2 + 2\,\xi\,M_{\bar D}\,q^0,
	\end{split}
	\label{eq:12}
\end{equation}
with $q^0$ the energy of the $K^+$ in the rest frame of the cluster
\begin{equation}\label{eq:13}
	 q^0=\frac{s-M_{K^+}^2-M_c^2}{2\,M_c},
\end{equation}
and
\begin{equation}\label{eq:14}
	\xi=\frac{M_c}{M_{D^*} + M_{\bar D}},
\end{equation}
which corresponds to distributing the binding energy between the $D^*$ and $\bar D$ in proportion to their masses.

The sum of the $ \tilde{T}'_{ij}$ matrices over $i,\,j$, which represents the total final amplitude for the $K^+X(3872)$ scattering ($D^*\bar D$ component for the moment), can be reduced to a compact formula shown in Ref.~\cite{Agatao:2025ckp}
\begin{equation}\label{eq:Ttot12}
	T_{ D^*\bar D}^{\,\mathrm{tot}}=\frac{
		\tilde{t}_1+\tilde{t}_2+\left(2\, G_0-G_c^{(1)}-G_c^{(2)}\right) \tilde{t}_1 \,\tilde{t}_2}
	{1-G_c^{(1)} \,\tilde{t}_1-G_c^{(2)} \,\tilde{t}_2-\left(G_0^2-G_c^{(1)} \,G_c^{(2)}\right) \tilde{t}_1 \,\tilde{t}_2}.
\end{equation}
The amplitudes $t_1,\,t_2$ are evaluated in the Appendix.

In order to take into account the $D \bar D^* $ component of the $X(3872)$, we note that the transition from $D^*$ to $\bar D^*$ requires at least two steps involving the exchange of charmed mesons and is therefore highly suppressed~\cite{Dias:2021upl}.
Consequently, the total amplitude is~\cite{Jia:2026dpl}
\begin{equation}\label{eq:Ttot_1234}
	(T^{\,\rm tot})^{-1}=\frac{1}{2}\Big(
	(T^{\,\mathrm{tot}}_{D\bar D^*})^{-1}+(T^{\,\mathrm{tot}}_{D^* \bar D})^{-1}
	\Big),
\end{equation}
\color{black}
\noindent where $T^{\,\mathrm{tot}}_{D \bar D^*}$ is evaluated as $T^{\,\mathrm{tot}}_{D^* \bar D }$ done above replacing $D^*\to D$; $\bar D\to \bar D^* $. These amplitudes are further discussed in the Appendix.
{
In Eq.~\eqref{eq:Ttot_1234} we average the inverse of the amplitudes to preserve elastic unitarity. In practice there is not much difference if one uses the average of the amplitudes, but preserving the elastic unitarity is one of the priorities in the present work.
}

\subsection{Scattering length and effective range}
Our amplitude $T^{\,\mathrm{tot}}$ can be related to the standard one of Quantum Mechanics via
\begin{equation}
	\begin{aligned}[b]
		-8\pi\sqrt{s} \left(T^{\,\mathrm{tot}}\right)^{-1} &= (f^{\mathrm{QM}})^{-1} \\[1mm]
		&\approx -\frac{1}{a} + \frac{1}{2}\,r_0 \,q_{\mathrm{cm}}^2 - i \,q_{\mathrm{cm}},
	\end{aligned}
	\label{eq:17}
\end{equation}
{where $q_{\mathrm{cm}}$  is the CM momentum of the system $K^+X(3872)$,}
with $-i\, q_{\mathrm{cm}}$ standing for elastic unitarity. 
As shown in Ref.~\cite{Agatao:2025ckp}, the amplitude $T^{\,\mathrm{tot}}$ satisfies exactly elastic unitarity and we can write
\begin{align}
	a &= {\frac{T^{\,\mathrm{tot}}~}{8\pi\sqrt{s}}\Big|_{\mathrm{th}}}, \label{eq:18} \\[2mm]
	r_0 &= \frac{1}{\mu} \left[ \frac{\partial}{\partial \sqrt{s}} \left( -8\pi\sqrt{s} \left(T^{\,\mathrm{tot}}\right)^{-1} + i\,q_{\mathrm{cm}} \right) \right]_{\mathrm{th}}, 
    \label{eq:19}
\end{align}
with $\mu$ the $K^+X(3872)$ reduced masss, and $th$ denoting the threshold.

{{
We find it useful to clarify the concept of elastic unitarity. Even when we have many coupled channels, with some of them open at the threshold of $K^+X(3872)$, which render the $K^+X$ amplitude complex at threshold, one can still define a complex optical potential 
$V(\sqrt{s})$, such that 
\begin{equation}\label{topt}
	\begin{aligned}[b]
		T=\frac{V_\text{opt}}{1-V_\text{opt}\,G_{KX}};\qquad T^{-1}=V_\text{opt}^{-1}-G_{KX}.
	\end{aligned}
\end{equation}
Taking into account that around the $K^+X$ threshold one has
\begin{equation}
	\begin{aligned}[b]
		\sqrt{s} = m_{K^+}+m_{X}+\frac{q_{\mathrm{cm}}^2}{2\mu} ,
	\end{aligned}
\end{equation}
one can expand $T^{-1}$ above the $K^+X$ threshold in powers of $q_\text{cm}$. Since $V_\text{opt}$ is a function of $\sqrt s$, the expansion of  $V_\text{opt}^{-1}$ gives powers of $q_\text{cm}^2$. So does the expansion of $\text{Re}~G_{KX}$. However,  $\text{Im}~G_{KX}$, which comes when in the propagation of ${K^+X}$ in intermediate states the two particles are placed on shell, is linear in $q_\text{cm}$. Actually, from Eqs.~(\ref{eq:17}) and (\ref{topt})  one has
\begin{equation}
	\begin{aligned}[b]
		-8\pi\sqrt{s}\,(-i) \,\text{Im}\,G_{KX}&= 8\pi\sqrt{s} \frac{-i}{8\pi\sqrt{s}}\,q_{\mathrm{cm}} = - i \,q_{\mathrm{cm}}.
	\end{aligned}
	\label{liearterm_eq}
\end{equation}
This means that $a$, $r_0$ can be complex, but the linear term in $q_\text{cm}$ in Eq.(~\ref{eq:17}) is always  $-i~q_\text{cm}$, coming from the elastic propagation of the initial-final state in the intermediate steps of the Lippmann-Schwinger series.
}}

\subsection{Correlation function}
The $K^+X(3872)$ correlation function is then written as
\begin{equation}
	\begin{aligned}[b]
		C_{KX}(p) = & ~1 + 4\pi \int_{0}^{\infty} \mathrm{d}r \, r^2 \,S_{12}(r) 
		\times \left\{ \left|j_0(pr) + TG\right|^2 - j_0^2(pr) \right\},
	\end{aligned}
	\label{eq:20}
\end{equation}
where{{~\cite{Encarnacion:2026zas}
\begin{equation}
	TG = T^\text{tot} \frac{1}{2} \Bigg( G_1(\sqrt{s},r) + G_2(\sqrt{s},r) \Bigg),
	\label{eq:21}
\end{equation}
}}
and $S_{12}(r)$ is the source function
\begin{equation}
	S_{12}(r)=\frac{1}{\left(
		4\pi R^2
		\right)^{3/2}}\;e^{-r^2/4R^2},
	\label{eq:22}
\end{equation}
with $R$ the radius of the source, where ($i=1,\,2$)
\begin{equation}
		\begin{aligned}[b]
		G_i(\sqrt{s},r) &= \int \frac{\mathrm{d}^3 q}{(2\pi)^3} \,
		\frac{ j_0(qr)\,F_c^{(i)}(q) }{\sqrt{s} - \omega_{K}(\vec q\,) - \omega_c(\vec q\,) + i\epsilon} \\[1.5mm]
		&\quad \times \dfrac{1}{2\,\omega_{K}(\vec q\,)}\, \dfrac{1}{2\,\omega_{c}(\vec q\,)}\;\Theta\left(q_{\mathrm{max}}^{(i)} - q_i^*\right).
	\end{aligned}
	\label{eq:23}
\end{equation}


\section{Results}\label{sec:res}
\begin{figure}[tb]
    \centering
    \includegraphics[width=0.48\textwidth]{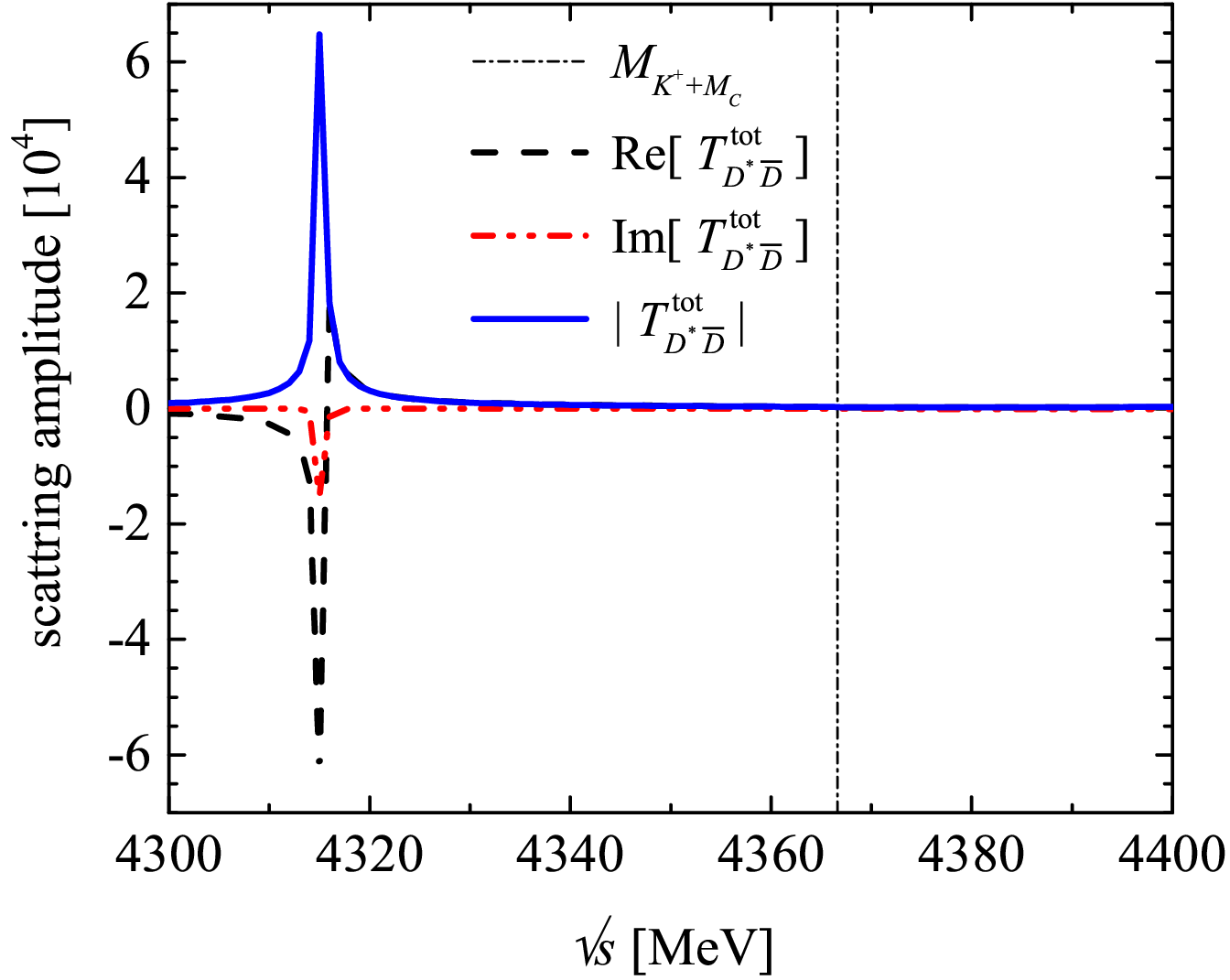}
    \caption{Results for \(T_{D^*\bar D}^{\mathrm{tot}}\) from Eq.~( \ref{eq:Ttot12}). The dashed line represents the real part, the dashed-dotted line the imaginary part and the solid line the modulus of the scattering amplitude. The vertical line corresponds to the threshold mass, \(M_{K^+}+M_X\).}
    \label{fig:T_DD}
\end{figure}

We first show the results for the three-body scattering amplitude \(T_{D^*\bar D}^{\,\mathrm{tot}}\) for the \(D^*\bar D\) component in Fig.~\ref{fig:T_DD}.
One can see that the amplitude exhibits a clear resonant structure below the threshold.
This behavior is characteristic of a bound or resonant state generated by the interaction.

\begin{figure}[tb]
    \centering
    \includegraphics[width=0.48\textwidth]{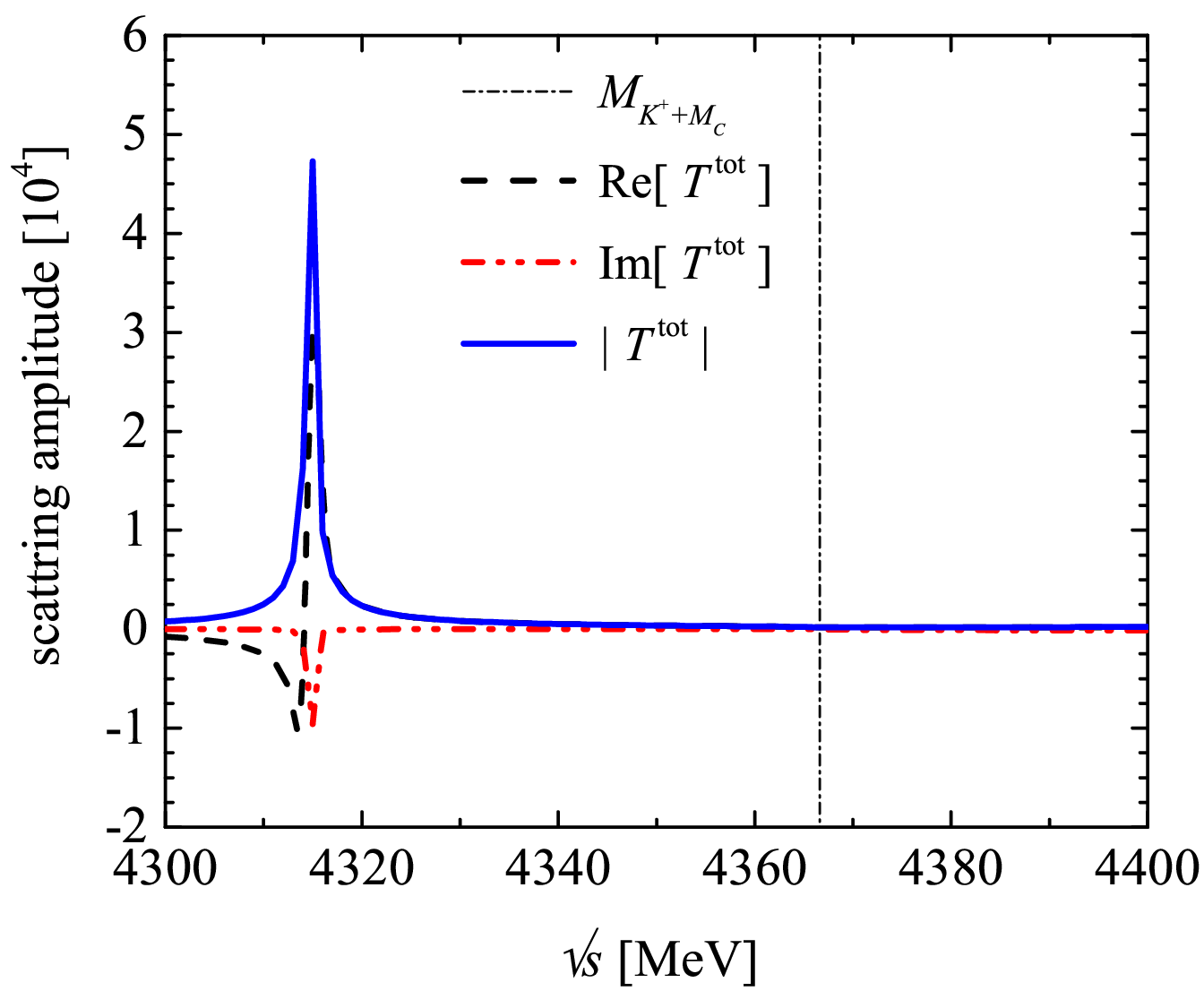}
    \caption{Result for \(T^{\mathrm{tot}}\) of Eq.~(\ref{eq:Ttot_1234}), considering the two components \(D^*\bar D\) and \(D\bar D^*\) of the \(X(3872)\) wave function.}
    \label{fig:T_tot}
\end{figure}

We then consider the full contribution from both molecular components of the \(X(3872)\) wave function, namely \(D^*\bar D\) and \(D\bar D^*\), given by Eq.~(\ref{eq:Ttot_1234}). The resulting total amplitude \(T^{\mathrm{tot}}\) is displayed in Fig.~\ref{fig:T_tot}. The behavior remains similar to that of the single-component case, confirming the robustness of the resonance-like structure.

The amplitude exhibits a clear resonant structure peaked around \(4315\,\mathrm{MeV}\), which is located about \(50\,\mathrm{MeV}\) below the \(K^+X(3872)\) threshold. The imaginary part reaches a maximum while the real part crosses zero at approximately the same energy, and the modulus \(|T^{\mathrm{tot}}|\) shows a neat and narrow peak. From the half-height width of Im\([ T^{\mathrm{tot}}]\), we estimate the width of the state to be around \(1\,\mathrm{MeV}\). This narrow width stems from the suppressed decay channels due to the molecular nature of the system. 
The attraction responsible for this peak below threshold originates
 from the $D_{s1}(2460)$ and $D_{s0}^*(2317)$ poles, which are 
dynamically generated in the $KD^*$ and $KD$ interaction, respectively, in $I = 0$.


Next, we evaluate the scattering length and effective range for the \(KX(3872)\) interaction. From the low-energy expansion of the amplitude, we obtain
\begin{align}
    a &= (0.39 - i\,0.00)\,\mathrm{fm}, \label{eq:a} \\
    r_0 &= (1.16 - i\,1.66)\,\mathrm{fm}. \label{eq:r0}
\end{align}
The positive real part of the scattering length is a consequence of the pole below threshold, while the imaginary part is essentially zero. 
This tiny imaginary part is consistent with the weak inelasticity in the $KX(3872)$ channel. The effective range has a significant negative imaginary part, which reflects the presence of nearby coupled-channel effects.

\begin{figure}[tb]
    \centering
    \includegraphics[width=0.48\textwidth]{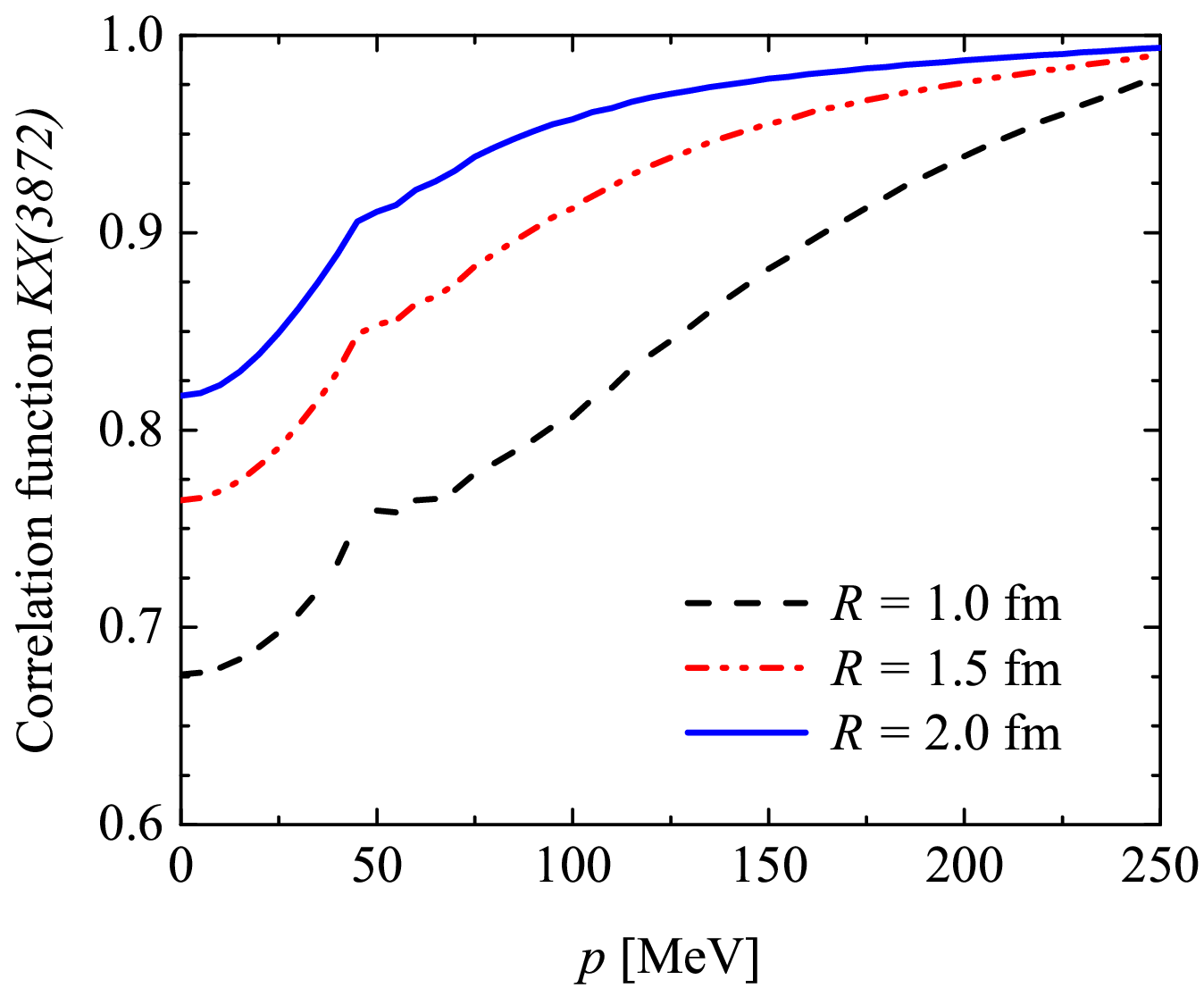}
    \caption{Correlation function for the \(KX(3872)\) system, considering the two components \(D^*\bar D\) and \(D\bar D^*\) of the \(X(3872)\) wave function.}
    \label{fig:corr}
\end{figure}

Finally, we compute the correlation function for the \(KX(3872)\) system, shown in Fig.~\ref{fig:corr}.
The correlation function deviates significantly from unity at low relative momenta, exhibiting a characteristic enhancement typical of a strongly attractive interaction that leads to a bound state. In the qualitative picture for correlation functions discussed in Refs.~\cite{Liu:2023uly,Liu:2024uxn}, this behavior corresponds to the case of a strongly attractive potential. The shape is very similar to those found in previous studies of analogous molecular systems, such as the \(p f_1(1285)\) in Ref.~\cite{Encarnacion:2025lyf} and the \(K f_1(1285)\) in Ref.~\cite{Jia:2026dpl}, all of which produce bound states with comparable binding energies. The shape is also similar to those obtained for the \(D^0K^+\) and \(D^+K^0\) correlation functions in Ref.~\cite{Ikeno:2023ojl}. One finds a small decrease around \(\vec{p}=50\,\mathrm{MeV}/c\) and no change at high momenta, consistent with the behavior expected from a  bound state generated by a strong attraction.



\section{Conclusions}\label{sec:concl}

We have investigated the interaction of a kaon with the \(X(3872)\) resonance within a unitarized fixed-center approximation, assuming the \(X(3872)\) to be a molecular state of \(D\bar D^*\) and \(D^*\bar D\) with isospin \(I=0\). In this framework, the \(X(3872)\) is treated as a cluster, while the kaon interacts with its \(D\) and \(D^*\) constituents. The optical potential obtained from the FCA is then used in a Lippmann–Schwinger equation to implement elastic unitarity in the \(KX(3872)\) channel, which is essential for the reliable evaluation of low-energy scattering observables.

We have determined the threshold scattering parameters for the \(KX(3872)\) system and obtained a scattering length \(a = (0.39 - i\,0.00)\,\mathrm{fm}\) and an effective range \(r_0 = (1.16 - i\,1.66)\,\mathrm{fm}\). The positive real part of \(a\) reflects the presence of a bound state below threshold.
{The  negligible imaginary part is originated from the $KD$ and $KD^*$ amplitudes.} The significant negative imaginary part of \(r_0\) indicates the presence of coupled-channel effects.

The \(KX(3872)\) scattering amplitude develops a neat and narrow resonant structure peaked around \(4315\,\mathrm{MeV}\), which lies about \(50\,\mathrm{MeV}\) below the \(K^+X(3872)\) threshold. From the half-height width of the imaginary part, we estimate the width of this three-body state to be approximately \(1\,\mathrm{MeV}\). This state is a direct consequence of the combined \(KD\) and \(K D^*\) attractions and its existence is intimately connected to the composite nature of the \(X(3872)\).

The correlation function of the \(KX(3872)\) system exhibits a characteristic enhancement at low relative momenta, consistent with a strongly attractive interaction leading to a bound state. Its shape is qualitatively similar to those found in related molecular systems such as \(p f_1(1285)\) and \(K f_1(1285)\), further supporting the robustness of our predictions. A small decrease around \(\vec{p}=50\,\mathrm{MeV}/c\) and no change at high momenta are observed, which is consistent with the behavior expected from the presence of a  bound state.


The observables derived in this work can also be employed in the inverse method of Refs.~\cite{Ikeno:2023ojl,Albaladejo:2023wmv} to extract the interaction parameters directly from future experimental data, providing a critical assessment of the molecular picture of the \(X(3872)\) and the validity of the three-body theoretical framework. We anticipate that searches for the predicted state, along with similar investigations of other three-body systems, will offer new insights into the nature of exotic hadrons and the dynamics of hadronic molecules.

\appendix
\section{Amplitudes $t_1$ and $t_2$}\label{app:amp}
In this appendix we detail the two-body scattering matrices $t_1$ and $t_2$ that enter the FCA equations of Sec.~\ref{sec:form}.
They correspond, respectively, to the interaction of the external kaon with the $D^*$ and $\bar D$ constituents of the $X(3872)$ cluster for the $D^*\bar D$ configuration, and analogously for the  $D\bar D^*$ configuration.

\subsection{The $K D^*\bar D$ system}
For the $D^*\bar D$ component of the $X(3872)$ we need the $K D^*$ and $K\bar D$ amplitudes in $I=0$ and $I=1$.
The isospin decomposition reads
\begin{align}
	t_1(\sqrt{s_{K D^*}}) &= \frac{3}{4}\,t_{K D^*}^{I=1} + \frac{1}{4}\,t_{K D^*}^{I=0}, \label{eq:app_t1a} \\
	t_2(\sqrt{s_{K\bar D}})   &= \frac{3}{4}\,t_{K\bar D}^{I=1}   + \frac{1}{4}\,t_{K\bar D}^{I=0},   \label{eq:app_t2a}
\end{align}
where the sub-energies $\sqrt{s_{K D^*}}$ and $\sqrt{s_{K\bar D}}$ are defined in Eqs.~\eqref{eq:12}--\eqref{eq:14} of the main text.

The $K D^*$ interaction in $I=0$ is dominated by the $D_{s1}(2460)$ resonance.
We use the Breit--Wigner parametrization
\begin{equation}
	t_{K D^*}^{I=0} = \frac{g_{D_{s1},K D^*}^2}{M^2_{\mathrm{inv}}(K D^*) - m_{D_{s1}}^2 + i\, m_{D_{s1}}\Gamma_{D_{s1}}},
	\label{eq:app_KDstarI0}
\end{equation}
with $m_{D_{s1}}=2460$~MeV, $\Gamma_{D_{s1}}=0.1$~MeV, and the coupling {$|g_{D_{s1},K D^*}|=12107$~MeV} taken from the chiral unitary approach of Ref.~\cite{Lin:2024hys}.
In the $I=1$ channel the interaction is null,
\begin{equation}
	t_{K D^*}^{I=1} = 0 .
	\label{eq:app_KDstarI1}
\end{equation}

For the $K\bar D$ interaction we employ the potentials derived from the local hidden gauge approach~\cite{Bando:1984ej,Bando:1987br,Meissner:1987ge,Nagahiro:2008cv} with the exchange of vector mesons.
Since the $I = 1$ potential is repulsive, the amplitude is taken directly as the Born term
\begin{equation}
	t_{K\bar D}^{I=1} = V_{K\bar D}^{I=1},
	\label{eq:app_KbarDI1}
\end{equation}
while the $I=0$ amplitude, corresponding to an attractive potential,  is unitarized via the Bethe--Salpeter equation
\begin{equation}
	t_{K\bar D}^{I=0} = \frac{1}{(V_{K\bar D}^{I=0})^{-1} - G},
	\label{eq:app_KbarDI0}
\end{equation}
where $G$ is the regularized loop function for the $K\bar D$ channel, given in Eq.~\eqref{eq:app_G} below.
After $S$-wave projection, the potentials read
\begin{align}
	V_{K\bar D}^{I=1} &=  \frac{g^2}{M_V^2}\,
		\frac{1}{2}\Bigg[ 3s - \big(M^2 + m^2 + M'^2 + m'^2 \big) \nonumber\\
		&\qquad\qquad -\frac{1}{s}\big(M^2 - m^2\big)\big(M'^2 - m'^2\big) \Bigg], \label{eq:app_VKbarDI1} \\[2mm]
	V_{K\bar D}^{I=0} &= -\frac{g^2}{M_V^2}\,
		\frac{1}{2}\Bigg[ 3s - \big(M^2 + m^2 + M'^2 + m'^2 \big) \nonumber\\
		&\qquad\qquad -\frac{1}{s}\big(M^2 - m^2\big)\big(M'^2 - m'^2\big) \Bigg], \label{eq:app_VKbarDI0}
\end{align}
with $g = M_V/(2f_\pi)$, $M_V=800$~MeV, $f_\pi=93$~MeV, and $M=M'=M_K$, $m=m'=M_{\bar D}$.
The loop function is regularized with a cutoff $q_{\max}=689$~MeV as in Ref.~\cite{Ikeno:2023ojl,Su:2025aiz},
\begin{equation}
	G(s) = \int_{|\vec q\,|<q_{\max}} \frac{\mathrm{d}^3 q}{(2\pi)^3}\,
		\frac{\omega_K(\vec q\,) + \omega_{\bar D}(\vec q\,)}{2\,\omega_K(\vec q\,)\,\omega_{\bar D}(\vec q\,)}\,
		\frac{1}{s - [\omega_K(\vec q\,) + \omega_{\bar D}(\vec q\,)]^2 + i\epsilon}.
	\label{eq:app_G}
\end{equation}

\subsection{The $K D\bar D^*$ system}
The treatment of the $D\bar D^*$ component follows the same strategy.
Now $t_1$ refers to the $K D$ interaction and $t_2$ to the $K \bar D^*$ interaction:
\begin{align}
	t_1(\sqrt{s_{K D}})      &= \frac{3}{4}\,t_{K D}^{I=1}      + \frac{1}{4}\,t_{K D}^{I=0},      \label{eq:app_t1b} \\
	t_2(\sqrt{s_{K\bar D^*}}) &= \frac{3}{4}\,t_{K\bar D^*}^{I=1} + \frac{1}{4}\,t_{K\bar D^*}^{I=0}. \label{eq:app_t2b}
\end{align}

The $KD$ amplitude in $I = 0$ dynamically generates 
the $D_{s0}^*(2317)$ resonance, and we take

\begin{equation}
	t_{K D}^{I=0} = \frac{g_{D_{s0},K D}^2}{M^2_{\mathrm{inv}}(K D) - m_{D_{s0}}^2 + i\, m_{D_{s0}}\Gamma_{D_{s0}}},
	\label{eq:app_KDI0}
\end{equation}
with $m_{D_{s0}}=2317$~MeV, $\Gamma_{D_{s0}}\simeq0.1$~MeV.
For the coupling  $g_{D_{s0},K D}$ we take 9600 MeV, considering the results obtained in Refs.~\cite{MartinezTorres:2014kpc,Ikeno:2023ojl}.
Once again, $t_{K D}^{I=1}=0$.

The $K\bar D^*$ amplitudes are built from the same vector-exchange potentials as before,
\begin{align}
	V_{K\bar D^*}^{I=1} &=  \frac{g^2}{M_V^2}\,
		\frac{1}{2}\Bigg[ 3s - \big(M^2 + m^2 + M'^2 + m'^2 \big) \nonumber\\
		&\qquad\qquad -\frac{1}{s}\big(M^2 - m^2\big)\big(M'^2 - m'^2\big) \Bigg], \\
	V_{K\bar D^*}^{I=0} &= -\frac{g^2}{M_V^2}\,
		\frac{1}{2}\Bigg[ 3s - \big(M^2 + m^2 + M'^2 + m'^2 \big) \nonumber\\
		&\qquad\qquad -\frac{1}{s}\big(M^2 - m^2\big)\big(M'^2 - m'^2\big) \Bigg],
\end{align}
with $M=M'=M_K$, $m=m'=M_{\bar D^*}$.
The unitarized $I=0$ amplitude is
\begin{equation}
	t_{K\bar D^*}^{I=0} = \frac{1}{(V_{K\bar D^*}^{I=0})^{-1} - G},
	\label{eq:app_KbarDstarI0}
\end{equation}
while $t_{K\bar D^*}^{I=1} = V_{K\bar D^*}^{I=1}$.
The loop function $G$ is the same as in Eq.~\eqref{eq:app_G} with the obvious replacement $\bar D\to\bar D^*$ and the same cutoff $q_{\max}=689$~MeV~\cite{Ikeno:2023ojl,Su:2025aiz}.

The amplitudes $t_1$ and $t_2$ constructed above are then normalized according to Eq.~\eqref{eq:1} of the main text to be used in the FCA formalism.

\section*{acknowledgments}
This work is partly supported by the National Natural Science
Foundation of China under Grants  No. 12405089 and
the China Postdoctoral Science Foundation under Grant
No. 2022M720360 and No. 2022M720359. J. Song acknowledges support from the Hainan Provincial Excellent Talent Team under the “Four Talents” Gathering Program of Hainan Province.
We also acknowledge support from the programs Unidades de
Excelencia Severo Ochoa CEX2023-001292-S and María de
Maeztu CEX2020-001058-M, and from the projects PID2022-
139427NB-I00 and PID2023-147458NB-C21 financed by the Spanish
MCIN/AEI/10.13039/501100011033/FEDER, UE (FSE+), and by the
Grant CIPROM 2023/59 of Generalitat Valenciana. P.B  gratefully acknowledges the partial support provided by the Coordena\c{c}\~{a}o de
Aperfei\c{c}oamento de Pessoal de N\'{\i}vel Superior -- Brasil
(CAPES) -- Finance Code 001.

\bibliography{refs.bib} 
\end{document}